\documentclass{PoS}

\title{Fuzzy spaces from tensor models, cyclicity condition, and n-ary algebras}

\ShortTitle{Fuzzy spaces from tensor models, cyclicity condition, and n-ary algebras}

\author{\speaker{Naoki Sasakura}%

         \\

        Yukawa Institute for Theoretical Physics, Kyoto University,
Kyoto 606-8502, Japan\\

        E-mail: \email{sasakura@yukawa.kyoto-u.ac.jp}}

\abstract{The rank-three tensor models, which have a rank-three tensor 
as their only dynamical variable, may be interpreted as models of
dynamical fuzzy spaces. In this interpretation,
the generalized Hermiticity condition on the rank-three tensor 
leads to a cyclic property of the algebra of functions on fuzzy spaces.
The fuzzy spaces with the cyclic property are shown to have 
various physically interesting characteristics.
(i) Although the function algebras of the kind are nonassociative in general, various properties analogous
to quantum mechanics hold on the fuzzy spaces. (ii) The symmetry of the rank-three tensor models can be shown to be 
represented systematically by $n$-ary transformations on the fuzzy spaces.
 The transformations contain, for instance, diffeomorphism on fuzzy spaces.
(iii) There exists a systematic procedure of truncating the function algebras of the kind, and it can be used to 
consider subspaces, compactifications, lattice theories, and coarse-graining procedures of fuzzy spaces 
in physical applications. 
}

\FullConference{Proceedings of the Corfu Summer Institute 2011 
"School and Workshops on Elementary Particle Physics and Gravity"\\

                 September 4-18, 2011\\

                 Corfu, Greece}

\begin{document}

\section{Introduction}
\label{sec:introduction}
The rank-three tensor models were originally introduced as models for 
the three-dimensional simplicial quantum gravity 
\cite{Ambjorn:1990ge,Sasakura:1990fs,Godfrey:1990dt}.
The dynamical variable of the rank-three tensor models is a rank-three tensor,
$M_{abc}\ (a,b,c=1,2,\ldots,N)$,
which satisfies the generalized Hermiticity condition,
\begin{equation}
M_{abc}=M_{bca}=M_{cab}=M_{bac}^*=M_{acb}^*=M_{cba}^*.
\label{eq:genher}
\end{equation} 
The symmetry of the rank-three tensor models is the orthogonal group symmetry,
\begin{equation}
M_{abc}'= O_{aa'}O_{bb'}O_{cc'}M_{a'b'c'}, \ \ O\in O(N).
\label{eq:orthogonal}
\end{equation}

The relation of the tensor models to the simplicial quantum gravity was given by the correspondence 
between the Feynman diagrams of the tensor models and the diagrams dual to the simplicial 
manifolds \cite{Ambjorn:1990ge,Sasakura:1990fs,Godfrey:1990dt}.
In this correspondence, the ranks of the tensor variables are directly related to the dimensions of the simplicial quantum gravity. 
This seems to be an unfavored property of the tensor models as quantum gravity, 
since the dimensions of spaces should be dynamical quantities rather than input parameters.
Moreover, since one has to take the limit of infinite numbers of vertices of the Feynman diagrams
to take the continuum limit of the simplicial quantum gravity, the tensor models must be computed non-perturbatively
to obtain physical results of quantum gravity. There have recently been
some major developments in this direction \cite{Gurau:2011xp,Oriti:2011jm}, but they are still limited. 
 
These difficulties may be circumvented by regarding the tensor models from another perspective.
The rank-three tensor models, which have a rank-three tensor as their only dynamical
variable, may be interpreted as models of dynamical fuzzy spaces 
\cite{Sasakura:2005js,Sasakura:2011ma}.
Since fuzzy spaces can in principle approximate any dimensional spaces, the rank-three tensor models 
should be able to describe any dimensional spaces.
Moreover, semi-classical treatments of the rank-three tensor models can be interpreted
physically; classical solutions correspond to emergent background fuzzy spaces, and 
the perturbations around them to emergent field theories on the background fuzzy spaces.
In fact, numerical semi-classical studies have shown emergent spaces \cite{Sasakura:2006pq} and emergent (Euclidean) general relativity
for a few fine-tuned rank-three tensor models \cite{Sasakura:2008pe,Sasakura:2009hs}.     

While a classical space is described by a manifold, 
a fuzzy space is described by an algebra of functions on it as
\begin{equation}
\phi_a \phi_b=f_{ab}{}^c \phi_c,
\label{eq:fuzzyalg}
\end{equation}
where $\{\phi_a\,|\,a=1,2,\ldots,N\}$ are the bases of functions on a fuzzy space, and 
$f_{ab}{}^c$ are the structure constants which define the function algebra.
One can consider various fuzzy spaces by changing the values of $f_{ab}{}^c$.
Noncommutative associative algebras define noncommutative spaces \cite{Connes:1994yd,Madore:1991bw}, 
while one can also get nonassociative spaces \cite{deMedeiros:2004wb,Ramgoolam:2003cs,Ramgoolam:2001zx,Sasai:2006ua}.
Another ingredient of the fuzzy spaces in this paper is the inner product \cite{Sasakura:2011ma},
\begin{equation}
\langle \phi_a | \phi_b \rangle=g_{ab},
\label{eq:inner}
\end{equation}
which is assumed to be bi-linear, symmetric and real. 

The correspondence between the rank-three tensor models and the fuzzy spaces is assumed to be
given by the following relation between the dynamical variable of the rank-three tensor models and 
the parameters of the fuzzy spaces as
\begin{equation}
M_{abc}=f_{ab}{}^{c'}g_{c'c}.
\label{eq:mfrelation}
\end{equation}
Then the generalized Hermiticity condition (\ref{eq:genher}) can be converted to the following
properties of the fuzzy spaces, 
\begin{eqnarray}
&&\langle \phi_a \phi_b |\phi_c\rangle=\langle \phi_a | \phi_b \phi_c \rangle
=\langle \phi_b | \phi_c \phi_a \rangle,\label{eq:cyclic}\\
&&(\phi_a \phi_b)^*=\phi_b \phi_a,
\label{eq:complexconj}
\end{eqnarray}
where $*$ denotes complex conjugation, and the functions are assumed to be real,
\begin{equation}
\phi_a^*=\phi_a.
\label{eq:phireal}
\end{equation}
The equation (\ref{eq:cyclic}) is a cyclic property of the algebra of 
the functions, which will play essential roles in the following discussions. 

To balance the degrees of freedom of the rank-three tensor models with those of the fuzzy spaces, 
the inner product is assumed to be fixed by 
\begin{equation}
g_{ab}=\delta_{ab}.
\label{eq:gfix}
\end{equation}
This is always possible if 
the matrix $g_{ab}$ in (\ref{eq:inner}) is positive definite, since it can be transformed to the form (\ref{eq:gfix})
by a $GL(N,R)$ transformation of the basis functions $\{\phi_a|a=1,2,\ldots,N\}$.  
Then the orthogonal symmetry (\ref{eq:orthogonal}) of the rank-three tensor models can be identified with the remaining
symmetry of the fuzzy spaces after the gauge-fixing (\ref{eq:gfix}).

\section{Quantum mechanical properties}
Associativity is a main feature of the algebras in quantum mechanics. 
However,  
the algebra (\ref{eq:fuzzyalg}) is not associative in general.  
Therefore there arises a doubt whether the kind of fuzzy spaces associated to the rank-three
tensor models are physically sensible. 
In this section,  
it will be shown that,
essentially due to the cyclicity property (\ref{eq:cyclic}), 
the fuzzy spaces indeed have various properties 
in common with quantum mechanics.

Let me first define some basic objects.
A state is defined by
\begin{equation}
|s\rangle=s_a |\phi_a\rangle,\ \langle s|=s_a \langle \phi_a |,
\label{eq:state}
\end{equation}
where $s_a$ are complex numbers in general. Note that, 
to respect the bilinear property of the inner product (\ref{eq:inner}),
the definition of $\langle s|$ in (\ref{eq:state}) 
does not contain taking the complex conjugate of $s_a$. One may instead 
define the usual bra-state of quantum mechanics by 
\begin{equation}
\langle\! \langle s| \equiv \langle s^* |=
s_a^* \langle \phi_a |.
\label{eq:brastate}
\end{equation}

An operator is defined by an expression, 
\begin{equation}
{\cal O}=v_a \phi_a,
\label{eq:operator}
\end{equation}
where $v_a$ are complex numbers in general, and 
is supposed to operate on a state as
\begin{equation}
{\cal O}|s\rangle=|{\cal O}s\rangle=v_a s_b |\phi_a \phi_b\rangle=v_a s_b f_{ab}{}^c |\phi_c \rangle.
\label{eq:operation}
\end{equation}
It is interesting to see the uniqueness of the matrix elements of the 
operator (\ref{eq:operator}). 
Since one can show from the cyclic property
(\ref{eq:cyclic}) and (\ref{eq:operation}) that
\begin{equation}
\langle \phi_a | {\cal O} | \phi_b \rangle=
\langle \phi_a | {\cal O}  \phi_b \rangle=v_c \langle \phi_a |\phi_c  \phi_b \rangle =
v_c \langle \phi_a \phi_c|  \phi_b \rangle =\langle \phi_a {\cal O}|\phi_b \rangle,
\label{eq:shift}
\end{equation}
the matrix elements can uniquely be defined in either way as 
$\langle \phi_a | {\cal O}|  \phi_b \rangle = \langle \phi_a | {\cal O}  \phi_b \rangle
=\langle \phi_a  {\cal O} | \phi_b \rangle$.

The matrix elements can compute ordered products of the operators.
One can show from the cyclic property (\ref{eq:cyclic}) that
\begin{eqnarray}
(O_1 O_2\cdots O_n)_{ab}&=&\langle\phi_a|{\cal O}_1({\cal O}_2 (\cdots ({\cal O}_n \phi_b))\cdots)
\rangle \nonumber \\
&=&\langle ((\cdots (\phi_a {\cal O}_1) {\cal O}_2)\cdots ){\cal O}_{p-1}|
{\cal O}_p({\cal O}_{p+1}(\cdots({\cal O}_n \phi_b))\cdots )\rangle\\
&=& \langle ((\cdots (\phi_a {\cal O}_1){\cal O}_2)\cdots){\cal O}_n | \phi_b \rangle,
\nonumber
\end{eqnarray}
where $O_i$ are the matrices which have the matrix elements $(O_i)_{ab}=\langle \phi_a |
{\cal O}_i |\phi_b \rangle$.

The Hermitian conjugate of an operator can be discussed as follows. One can show from
(\ref{eq:complexconj}),  
(\ref{eq:state}), (\ref{eq:brastate}) and (\ref{eq:shift}) that
\begin{equation}
\langle \!\langle s_1 | {\cal O} s_2 \rangle=\langle s_1^* |{\cal O} s_2\rangle
=\langle s_1^* {\cal O} | s_2\rangle
=\langle\! \langle {\cal O}^* s_1 | s_2 \rangle.
\end{equation} 
Therefore the Hermitian conjugate of an operator ${\cal O}$ is given by its complex 
conjugate ${\cal O}^*$. 

Then an observable can be defined by ${\cal O}={\cal O}^*$.
In fact, the mean value of an operator satisfying ${\cal O}={\cal O}^*$ is real as
\begin{equation}
\langle\!\langle s|{\cal O}|s\rangle^*=
\langle \!\langle s|{\cal O}  s\rangle^*=
\langle \!\langle {\cal O} s | s  \rangle=\langle\! \langle s | {\cal O} s \rangle
=\langle \! \langle s | {\cal O}| s \rangle.
\end{equation}
In  general, one can show for ${\cal O}={\cal O}^*$ that 
\begin{equation}
\langle \!\langle \phi_a | {\cal O} | \phi_b \rangle^*=\langle\! \langle \phi_a | {\cal O} \phi_b \rangle^*
=\langle \!\langle {\cal O} \phi_b | \phi_a \rangle=\langle \!\langle \phi_b | {\cal O} \phi_a \rangle
=\langle \!\langle \phi_b | {\cal O} | \phi_a \rangle.
\end{equation}
This means that the matrix $\langle\! \langle \phi_a | {\cal O} | \phi_b \rangle$ is 
hermite and can be transformed by a unitary matrix to a real diagonal matrix. 
The unitary matrix can be used to define the eigenstates, 
$|s_i\rangle=u_{ia}|\phi_a\rangle,\ (i=1,2,\ldots,N)$, which satisfy
\begin{equation}
{\cal O}|s_i\rangle =e_i |s_i\rangle,\ \langle\!\langle s_i | s_j \rangle=\delta_{ij},\ e_i\hbox{\,: real}.
\end{equation}
Thus an operator satisfying ${\cal O}={\cal O}^*$ can be diagonalized by the eigenstates with real eigenvalues, and 
can be qualified as an observable as in quantum mechanics.

Let me next discuss an uncertainty relation on the fuzzy spaces. Let me consider an observable 
${\cal O}={\cal O}^*$.
As discussed above, a real value is obtained as its mean value,
\begin{equation}
\langle {\cal O} \rangle \equiv \langle \!\langle s|{\cal O}|s\rangle,
\end{equation}
where the normalization $\langle\! \langle s | s \rangle =1$ is assumed.
Then the mean-square deviation can be defined by
\begin{eqnarray}
(\Delta {\cal O})^2&\equiv&\langle\! \langle ({\cal O}-\langle {\cal O} \rangle) s | 
({\cal O}-\langle {\cal O} \rangle) s \rangle \\
&=& \langle\! \langle{\cal O} s |{\cal O} s\rangle - \langle {\cal O}\rangle^2. \nonumber 
\end{eqnarray}
The following quantity is obviously positive, 
\begin{eqnarray}
&&\langle \!\langle ({\cal O}_1-\langle {\cal O}_1 \rangle +i\lambda ({\cal O}_2-\langle {\cal O}_2\rangle))s
|({\cal O}_1-\langle {\cal O}_1 \rangle +i\lambda ({\cal O}_2-\langle {\cal O}_2\rangle))s \rangle \nonumber \\
&&\ \ \ \ \ \ \ \ \ =(\Delta {\cal O}_1)^2 +i \lambda \langle \! \langle s| [{\cal O}_1,{\cal O}_2;s]\rangle+
\lambda^2 (\Delta {\cal O}_2)^2 \geq 0,
\label{eq:positive}
\end{eqnarray}
where ${\cal O}_i$ are observables, $\lambda$ is assumed to be real, and $[\ ,\ ;\ ]$ denotes the commutation
of two ordered operations,
\begin{equation}
[{\cal O}_1,{\cal O}_2;s]\equiv {\cal O}_1({\cal O}_2 s)- {\cal O}_2({\cal O}_1 s).
\label{eq:threebracket}
\end{equation} 
In the derivation of (\ref{eq:positive}), the cyclic property (\ref{eq:cyclic}) has played essential roles.
The fact that the inequality (\ref{eq:positive}) holds 
for any $\lambda$ leads to an inequality,
\begin{equation}
\Delta {\cal O}_1 \Delta {\cal O}_2 \geq \frac{1}{2} 
\left|\langle \! \langle s|[{\cal O}_1,{\cal O}_2;s]\rangle \right|. 
\label{eq:uncertainty}
\end{equation} 
This is the uncertainty relation which holds generally on the fuzzy spaces that can be associated to the rank-three tensor models.

The 3-bracket $[\ ,\ ;\ ]$, which is defined in (\ref{eq:threebracket}) and gives the lower bound for the 
uncertainty in (\ref{eq:uncertainty}), 
is a quantity affected by both the noncommutativity and the nonassociativity of an algebra.
Indeed, if the algebra is associative, 
\begin{equation}
[{\cal O}_1,{\cal O}_2;s]=[{\cal O}_1,{\cal O}_2]s.
\end{equation} 
Therefore, from (\ref{eq:uncertainty}), the uncertainty is bounded by the noncommutativity 
between ${\cal O}_1$ and ${\cal O}_2$, as  in quantum mechanics.
On the other hand, if the algebra is commutative, one obtains
\begin{equation}
[{\cal O}_1,{\cal O}_2;s]={\cal O}_1({\cal O}_2 s)- {\cal O}_2({\cal O}_1 s)
={\cal O}_1(s {\cal O}_2 )- ({\cal O}_1 s){\cal O}_2,
\end{equation}
which is the associator \cite{Okubo:1990nv} among ${\cal O}_i,s$.
Therefore the uncertainty is bounded by the nonassociativity of an algebra in this case.

Let me finally discuss the consequence, if a function algebra is 
neither noncommutative nor nonassociative, namely,
\begin{eqnarray}
&&[\phi_a,\phi_b]=0, 
\label{eq:commutative}\\
&&[\phi_a,\phi_b;\phi_c]=0,
\label{eq:associative} 
\end{eqnarray}
for any $\phi_a,\phi_b,\phi_c$. 
Then (\ref{eq:genher}), (\ref{eq:fuzzyalg}), (\ref{eq:mfrelation}) and (\ref{eq:commutative}) 
imply that $M_{abc}$ is totally symmetric with respect to the indices and real.
Furthermore, (\ref{eq:gfix}) and (\ref{eq:associative}) imply that the symmetric matrices $M_a$, defined by  
$(M_a)_{bc}\equiv M_{abc}$, are commutative among each other.
Therefore these matrices $M_a$ can simultaneously be diagonalized to
real diagonal matrices by using the 
orthogonal group symmetry (\ref{eq:orthogonal}). By taking also into account the total symmetry of the indices,
$M_{abc}$ is transformed to the totally diagonal form, $M_{abc}=m_a \delta_{ab}\delta_{bc}$ with
real $m_a$. This means that the function algebra can be transformed to the form,
\begin{equation}
\phi_a \phi_b=m_a \delta_{ab} \phi_a.
\end{equation}
This function algebra represents just a collection of independent points, but not 
a ``fuzzy'' space.

\section{Truncation of function algebras}
In physics, there exist various occasions in which one takes a subset of functions.
An example is to consider a subspace. 
In this case, two functions which take the same values on the subspace but may take
different values outside are considered to be equivalent.  
Therefore it is enough to take a part of them.   
The second example is a compactification of a space. In this case, one selects out functions
which satisfy a periodicity condition.
The third is to construct a lattice theory from a continuum theory. What is relevant
is the functional values only on lattice points, and the situation is similar to considering a subspace above.
The last one is a coarse graining procedure. In a coarse-graining procedure, one averages 
the values of dynamical variables over coarse-grained regions. This process may be regarded as 
choosing out slowly varying functions.
The theme of this section is to formulate the general truncation procedure for the function algebras
of the fuzzy spaces of the kind which can be associated to the rank-three tensor models.

By appropriately taking the basis of the functions $\{ \phi_a| a=1,2,\ldots,N\}$
for each purpose of problems in physics as illustrated in the preceding paragraph, 
a subspace of the functions can be considered by taking a subset of the basis functions,
\begin{equation}
\tilde A=\{\phi_a|a=1,2,\ldots,\tilde N\},\ \ \tilde N <N.
\end{equation}
For the subset, let me define 
\begin{eqnarray}
\tilde g_{ab}&\equiv& \langle \phi_a | \phi_b \rangle , \\
\tilde M_{abc}&\equiv&\langle \phi_a \phi_b | \phi_c \rangle,
\end{eqnarray}
where $a,b,c=1,2,\ldots,\tilde N$. Since $\tilde A$ is a part of the whole set, it is obvious that  
$\tilde g_{ab}$ is symmetric and real, and that $\tilde M_{abc}$ satisfies the generalized 
Hermiticity condition coming from (\ref{eq:genher}), as
\begin{eqnarray}
&&\tilde g_{ab}=\tilde g_{ba}=\tilde g_{ab}^*, \\ 
&&\tilde M_{abc}=\tilde M_{bca}=\tilde M_{cab}=\tilde M_{bac}^*=\tilde M_{acb}^*=\tilde M_{cba}^*,
\end{eqnarray}
where $a,b,c=1,2,\ldots,\tilde N$.
Let me assume that $\tilde g_{ab}$ is an invertible matrix. Then one can define 
a new algebra for the subset $\tilde A$ by
\begin{eqnarray}
\tilde \phi_a \tilde \phi_b &=&\tilde f_{ab}{}^c \tilde \phi_c \\
\tilde f_{ab}{}^c&\equiv & \tilde M_{abc'}\tilde g^{c'c},
\end{eqnarray}
where $\tilde g^{ab}$ is the inverse of $\tilde g_{ab}$. The inner product of $\tilde A$ is defined by 
\begin{equation}
\langle \tilde \phi_a | \tilde \phi_b \rangle = \tilde g_{ab},
\end{equation}
which is bi-linear, symmetric and real. It is obvious to prove
\begin{eqnarray}
&&\langle \tilde \phi_a  \tilde \phi_b | \tilde \phi_c\rangle=
\langle  \tilde \phi_a |  \tilde \phi_b  \tilde \phi_c \rangle
=\langle  \tilde \phi_b |  \tilde \phi_c  \tilde \phi_a \rangle,\label{eq:tildecyclic}\\
&&( \tilde \phi_a \tilde  \phi_b)^*= \tilde \phi_b  \tilde \phi_a,
\end{eqnarray}
where $\tilde \phi_a^*=\tilde \phi_a$.

A coarse-graining procedure of the fuzzy spaces may be considered as 
the following iterative procedure.
One first obtains a new algebraic structure 
by computing new structure constants 
$\bar f_{ab}{}^c$ and inner product $\bar g_{ab}$ from the original 
$f_{ab}{}^c$ and $g_{ab}$ through an algorithm\footnote{An application of a coarse-graining procedure to the
rank-three tensor models is discussed in \cite{Sasakura:2010rb}.}.
Then one selects out a part of the new algebra which is considered to be important for the dynamics of a physical system.
It would not be possible to define such a selection procedure in general, but it would 
be instructive to give an abstract example as follows. Let me assume 
that the matrix $\bar g_{ab}$ is positive definite, and define 
\begin{equation}
H_{ab}=\bar g^{cd} \langle \bar \phi_c \bar \phi_a | \bar \phi_b \bar \phi_d\rangle,
\end{equation}   
where $\bar g^{ab}$ is the inverse of $\bar g_{ab}$. It is easy to prove that  
the matrix $H_{ab}$ is an Hermitian matrix which is semi-positive definite.
Therefore a criterion to choose an important part would be to select out the directions
which take relatively large lengths with respect to the measure defined by $H_{ab}$.  

\section{$N$-ary transformations as the symmetry of the rank-three tensor models}
\label{sec:nary}
The symmetry of the rank-three tensor models is the orthogonal group symmetry (\ref{eq:orthogonal}).
This corresponds to the remaining symmetry of the fuzzy spaces after the gauge fixing (\ref{eq:gfix}),
as explained in the last paragraph of Section \ref{sec:introduction}. 
The purpose of this section is to show that the symmetry transformations can systematically be 
constructed by $n$-ary transformations \cite{FigueroaO'Farrill:2008bd,deAzcarraga:2010mr} on the fuzzy 
spaces \cite{Sasakura:2011ma,Sasakura:2011nj}. The cyclic property (\ref{eq:cyclic})
again plays essential roles.

Let me start with a simple example.
Let me define an infinitesimal transformation, 
\begin{equation}
\delta \phi_a=\phi_m(\phi_a \phi_n)-\phi_n(\phi_a \phi_m),
\label{eq:naryexample}
\end{equation}
where $m,n$ are considered to be the label of the transformation.
As explained in the last paragraph of Section \ref{sec:introduction}, 
the symmetry transformations of the rank-three tensor models can be identified with the transformations
of the functions 
which are real and keep invariant the inner product (\ref{eq:inner}) with (\ref{eq:gfix}).
Indeed one can show that the infinitesimal transformation (\ref{eq:naryexample})
conserves the inner product  as
\begin{eqnarray}
\delta \langle \phi_a| \phi_b \rangle &=& \langle \delta \phi_a| \phi_b \rangle
+\langle \phi_a| \delta \phi_b \rangle \nonumber \\
&=& \langle \phi_m (\phi_a \phi_n)|\phi_b \rangle +
\langle \phi_a | \phi_m (\phi_b \phi_n) \rangle -(m\leftrightarrow n) \nonumber \\
&=& \langle \phi_a |\phi_n(\phi_b \phi_m) \rangle +
\langle \phi_a | \phi_m (\phi_b \phi_n) \rangle -(m\leftrightarrow n) \nonumber \\
&=&0,
\end{eqnarray}
where I have used the cyclic property (\ref{eq:cyclic}) from the second to the third line.

The general procedure to construct an $n$-ary transformation which keeps invariant the 
inner product can be described as follows. Let me consider a product of 
$\phi_{m_1},\phi_{m_2},\ldots,\phi_{m_n},\phi_a$,
\begin{equation}
(\phi_{m_1},\phi_{m_2},\ldots,\phi_{m_n},s;\phi_a),
\end{equation}
where the label $s$ is an abstract notation which dictates
the order of the product of $\phi_{m_1},\phi_{m_2},\ldots,\phi_{m_n},\phi_a$.
By using the cyclic property (\ref{eq:cyclic}), one can always find a transpose product $\bar s$, which
satisfies 
\begin{equation}
\langle (\phi_{m_1},\phi_{m_2},\ldots, \phi_{m_n},s;\phi_a)|\phi_b \rangle=
\langle \phi_a|(\phi_{m_1},\phi_{m_2},\ldots, \phi_{m_n},\bar s;\phi_b) \rangle.
\end{equation}
Then one can easily show that the infinitesimal transformation,
\begin{equation}
\delta \phi_a=[\phi_{m_1},\phi_{m_2},\ldots, \phi_{m_n},s;\phi_a],
\label{eq:delnary}
\end{equation}
where the $n+1$-ary product is defined by
\begin{equation}
[\phi_{m_1},\phi_{m_2},\ldots, \phi_{m_n},s;\phi_a] \equiv
(\phi_{m_1},\phi_{m_2},\ldots, \phi_{m_n},s;\phi_a)-
(\phi_{m_1},\phi_{m_2},\ldots, \phi_{m_n},\bar s;\phi_a),
\end{equation}
keeps invariant the inner product as
\begin{eqnarray}
\delta \langle \phi_a | \phi_b \rangle &=& \langle \delta  \phi_a | \phi_b \rangle+ 
\langle \phi_a | \delta \phi_b \rangle \nonumber \\
&=& \langle [\phi_{m_1},\phi_{m_2},\ldots, \phi_{m_n},s;\phi_a] | \phi_b \rangle+
\langle \phi_a | [\phi_{m_1},\phi_{m_2},\ldots, \phi_{m_n},s;\phi_b] \rangle   \\
&=& \langle (\phi_{m_1},\phi_{m_2},\ldots, \phi_{m_n},s;\phi_a) | \phi_b \rangle
-\langle (\phi_{m_1},\phi_{m_2},\ldots, \phi_{m_n},\bar s;\phi_a) | \phi_b \rangle \nonumber \\
&&+\langle \phi_a | (\phi_{m_1},\phi_{m_2},\ldots, \phi_{m_n},s;\phi_b) \rangle 
-\langle \phi_a | (\phi_{m_1},\phi_{m_2},\ldots, \phi_{m_n},\bar s;\phi_b) \rangle
\nonumber \\
&=&0.\nonumber
\end{eqnarray}
To be consistent with the reality (\ref{eq:phireal}) of $\phi_a$, the 
infinitesimal transformation (\ref{eq:delnary}) must be made real,
for instance, by adding its complex conjugate. 

The results of this section can be generalized to the supersymmetric case \cite{Sasakura:2011qg}.

\section{Unbroken $n$-ary symmetry}
A fuzzy space may be considered by taking a non-vanishing background value of $M_{abc}$, which may be 
obtained as a classical solution of a rank-three tensor model.
The symmetry of the rank-three tensor model is spontaneously broken by the $M_{abc}$.
Let me suppose that there remains an unbroken symmetry.  
A generator $T_{ab}$ of the remaining symmetry satisfies
\begin{equation}
T_{aa'}M_{a'bc}+T_{bb'}M_{ab'c}+ T_{cc'}M_{abc'}=0,
\label{eq:condt}
\end{equation}
where $T_{ab}=-T_{ba}$.
In most cases, one can represent such an infinitesimal symmetry transformation by a linear combination
of $n$-ary transformations \cite{Sasakura:2011nj}.
One can indeed discuss such general cases of linear combinations, but for simplicity, in the following discussions, 
let me suppose that the generator can be represented by a single $n+1$-ary transformation,
\begin{equation}
T_{aa'} \phi_{a'}=[\phi_{m_1},\phi_{m_2},\ldots, \phi_{m_n},s;\phi_a],
\label{eq:naryrep}
\end{equation}
which is the kind discussed in Section \ref{sec:nary}.
From (\ref{eq:fuzzyalg}), (\ref{eq:mfrelation}), (\ref{eq:gfix}), (\ref{eq:condt}) and (\ref{eq:naryrep}), 
it is easy to prove the Leibnitz rule,
\begin{equation}
[\phi_{m_1},\phi_{m_2},\ldots, \phi_{m_n},s;\phi_a \phi_b]=
[\phi_{m_1},\phi_{m_2},\ldots, \phi_{m_n},s;\phi_a ]\phi_b
+\phi_a [\phi_{m_1},\phi_{m_2},\ldots, \phi_{m_n},s;\phi_b].
\label{eq:Leibnitzrule}
\end{equation}
Furthermore, the fundamental identity,
\begin{eqnarray}
[\phi_{m_1},\phi_{m_2},\ldots, \phi_{m_n},s; [\phi_{p_1},\phi_{p_2},\ldots, &&\phi_{p_{n'}},s';\phi_a]]=
[[\phi_{m_1},\phi_{m_2},\ldots, \phi_{m_n},s; \phi_{p_1}],\phi_{p_2},\ldots, \phi_{p_{n'}},s';\phi_a]
\nonumber \\
&&+[\phi_{p_1},[\phi_{m_1},\phi_{m_2},\ldots, \phi_{m_n},s; \phi_{p_2}],\ldots, \phi_{p_{n'}},s';\phi_a]
+\cdots \nonumber \\
&&+[\phi_{p_1},\phi_{p_2},\ldots, \phi_{p_{n'}},s';[\phi_{m_1},\phi_{m_2},\ldots, \phi_{m_n},s; \phi_a]],
\label{eq:fi}
\end{eqnarray}
holds, where 
$[\phi_{p_1},\phi_{p_2},\ldots, \phi_{p_{n'}},s'; \phi_a]$ is an arbitrary $n'+1$-ary product of the kind discussed 
in Section \ref{sec:nary},
since $[\phi_{p_1},\phi_{p_2},\ldots, \phi_{p_{n'}},s'; \phi_a]$ is a sum of
products to which the Leibnitz rule (\ref{eq:Leibnitzrule}) is applicable. 

If $[\phi_{p_1},\phi_{p_2},\ldots, \phi_{p_{n'}},s'; \phi_a]$ in (\ref{eq:fi}) is taken to be
an unbroken symmetry transformation with $n'=n$, the fundamental identity (\ref{eq:fi}) 
implies that the commutator of two $n+1$-ary transformations of an unbroken symmetry is given by 
a linear combination of $n+1$-ary transformations. 
This means that the unbroken symmetry is represented by a Lie $n+1$-algebra.

An example of an unbroken symmetry represented by $3$-ary transformations \cite{Sasakura:2011ma}
can be given 
for a fuzzy $D$-dimensional flat space. 
The algebra of functions is assumed to be given by \cite{Sasai:2006ua}
\begin{equation}
\phi_{p_1} \phi_{p_2}=\exp[-\alpha ((p_1)^2+(p_2)^2+(p_1+p_2)^2)] \phi_{p_1+p_2},
\label{eq:fuzzyflat}
\end{equation}  
where $p_i$ are $D$-dimensional momenta and $\alpha$ is a positive parameter. The algebra (\ref{eq:fuzzyflat}) is a 
nonassociative deformation of the algebra of the plane waves on a usual $D$-dimensional flat space.
The algebra (\ref{eq:fuzzyflat}) obviously respects the Poincare symmetry.

Let me define the ``coordinates'' of the fuzzy flat space by 
\begin{equation}
x^\mu\equiv \left. -i \frac{\partial\phi_p}{\partial p_{\mu}} \right| _{p=0}.
\label{eq:coordinates}
\end{equation}
This definition comes from an expected identification $\phi_p \sim e^{ipx}$. 
From explicit computations using (\ref{eq:fuzzyflat}) and (\ref{eq:coordinates}), one can show that
\begin{eqnarray}
[x^\mu,x^\nu;x^\rho]&=&4 \alpha (\delta^{\mu\rho}x^\nu-\delta^{\nu\rho}x^\mu), 
\label{eq:3aryrot}\\ 
\ [x^\mu,\phi_0;x^\nu]&=&4 \alpha \delta^{\mu\nu} \phi_0,
\label{eq:3arytrans}
\end{eqnarray}
where $\phi_0\equiv \phi_{p=0}$, and the 3-ary product is defined by the associator,
\begin{equation}
[\phi_a,\phi_b;\phi_c]\equiv (\phi_a \phi_c)\phi_b-\phi_a (\phi_c \phi_b).
\label{eq:3arycomm}
\end{equation}
One can see that (\ref{eq:3aryrot}) and (\ref{eq:3arytrans}) generate 
the rotations and the translations
of the Poincare symmetry, respectively. Therefore they form a Lie 3-algebra of the Poincare symmetry.

\section{Spontaneously broken $n$-ary symmetry}
\label{sec:spontaneous}
The gauge transformations are represented non-linearly with inhomogeneous terms.
Therefore the gauge symmetries and also the diffeomorphism symmetry as well have often been treated
as spontaneously broken symmetries \cite{Ferrari:1971at,Brandt:1974jw,Borisov:1974bn}.
In this section, the diffeomorphism will be discussed as $3$-ary transformations of spontaneously broken symmetry 
on the fuzzy flat space defined by the algebra (\ref{eq:fuzzyflat}).

The functions in (\ref{eq:fuzzyflat}) are labeled with momenta, but in the following discussions, 
it is more convenient to label them with coordinates.  By Fourier transformation of the momentum label, 
$\phi_x\equiv \int d^D p\, e^{ipx} \phi_p$, one obtains the function algebra in the coordinate representation as
\begin{equation}
\phi_{x}\phi_{y}=\int d^Dz\, \exp [-\beta ((x-y)^2+(x-z)^2+(y-z)^2)] \phi_z,
\label{eq:alginx}
\end{equation}   
where $\beta$ is a positive constant.
The usual space is obtained in the limit $\beta \rightarrow +\infty$. 
Let me consider the following infinitesimal transformation,
\begin{equation}
\delta \Phi=const. \int d^Dx\ [\phi_x,\phi_{x+\epsilon(x)};\Phi],
\label{eq:3arydiff}
\end{equation}
where the 3-ary product is defined in (\ref{eq:3arycomm}), and $\varepsilon(x)$ is an infinitesimal function of $x$.
An explicit computation using (\ref{eq:alginx}) shows that the infinitesimal 
transformation (\ref{eq:3arydiff}) generates
\begin{equation}
\delta \psi(x)=\epsilon^\mu(x)\partial_\mu \psi(x)+
\frac{1}{2}(\partial_\mu\epsilon^\mu(x))\psi(x)+O(\beta^{-1}),
\label{eq:deltav}
\end{equation}
where $\psi(x)$ is a function of $x$  defined by
\begin{equation}
\Phi=\int d^Dx\ \psi(x)\, \phi_x.
\label{eq:defv}
\end{equation} 
Physically, $\psi(x)$ is a field on the fuzzy space.

In the limit of the usual space $\beta \rightarrow + \infty$, the transformation (\ref{eq:deltav}) implies
that $\psi(x)$ is transformed as a scalar
{\it half} density rather than a scalar. This is consistent with the diffeomorphism symmetry.
If there are two scalar functions $h(x),f(x)$, the diffeomorphism invariant integration
over a space is given  by $\int d^Dx\, \sqrt{g(x)}\, h(x)f(x)$, where $g(x)$ is the determinant of 
the metric tensor. On the other hand, (\ref{eq:defv}) does not contain $\sqrt{g(x)}$,
but can be made diffeomorphism invariant by assuming that 
$\psi(x)$ and $\phi_x$ be transformed in the same manner as 
$g(x)^{1/4} h(x)$ and $g(x)^{1/4} f(x)$, which are scalar half densities.
Generally, the index contraction of the functions on the fuzzy flat space is assumed to be given by $h_a f_a=\int d^Dx\, h_x f_x$,
which is invariant under the transformation (\ref{eq:deltav}). 

\section{Scalar field action}
The algebraic framework presented in this paper can be applied to the construction of a scalar
field theory. Let me consider an action defined by
\begin{equation}
S=-\langle \Phi^* \phi_a | \phi_a \Phi \rangle+ m_0^2 \langle \Phi^*|\Phi\rangle, 
\label{eq:scalaraction}
\end{equation}
where $\Phi=\psi_a \phi_a$ with complex $\psi_a$. While the algebra of $\phi_a$ determines 
a background fuzzy space, $\psi_a$ can be regarded as a field on the fuzzy space. 
As discussed in Section \ref{sec:spontaneous},
the field is expected to become a scalar half density in the limit of a usual space. 

An advantage of expressing the action $S$ in the form (\ref{eq:scalaraction}) is that 
it is obviously invariant under the symmetry of the tensor models. Therefore,
if the background fuzzy space can well be identified with a usual space, 
the action $S$ should become a usual scalar field action which respects some symmetries. 
As for unbroken symmetries, the symmetries could be Poincare symmetries, spherical symmetries 
and/or supersymmetries, and as for spontaneously broken symmetries, 
the action should be invariant under diffeomorphism. Therefore,
if a fuzzy space which corresponds to a curved space is considered,
the action $S$ should reproduce 
a scalar field action on a curved space. 

One can obtain explicit forms of the action $S$ by specifying background fuzzy spaces. 
Let me first consider the fuzzy flat space. By substituting the algebra (\ref{eq:fuzzyflat})
into (\ref{eq:scalaraction}), one obtains
\begin{equation}
S_{flat}=\left(m_0^2-c_0 \exp (-3 \alpha p^2)\right) \psi_{p}^* \psi_{p},
\end{equation}
where $c_0$ is a positive number. In the low momentum region $\alpha p^2 \ll 1$, 
the dispersion can be approximated by
$(m_0^2-c_0)+3 \alpha c_0  p^2 +\cdots$, which is the standard kinetic term of
a scalar field on a flat space.

To give an example of a curved space, let me consider a two-dimensional fuzzy sphere.
It is convenient to label the functions by $(j,m)$, where $j=0,1,\ldots$, and $m=-j,-j+1,\ldots,j$.
The algebra of functions is assumed to be given by
\begin{equation}
\phi_{(j_1,m_1)} \phi_{(j_2,m_2)}=
\sum_{j_3,m_3} \frac{\prod_{i=1}^3 \sqrt{2j_i+1}D(j_i)}{\sqrt{4 \pi}} 
\left(
\begin{array}{ccc}
j_1 & j_2 & j_3 \\
0&0&0
\end{array}
\right)
\left(
\begin{array}{ccc}
j_1 & j_2 & j_3 \\
m_1&m_2&m_3
\end{array}
\right) (-1)^{m_3}
\phi_{(j_3, -m_3)},
\label{eq:spherealg}
\end{equation}
where $D(j)$ is a damping factor which vanishes 
in the limit $j \rightarrow \infty$.
If one substitutes the constant $D(j)=1$ into (\ref{eq:spherealg}), 
the algebra is that of the spherical harmonics on a two-sphere.
The damping factor $D(j)$ introduces fuzziness to the sphere by 
cutting off higher $j$ modes.
Note that the fuzzy space is a nonassociative two-sphere, since 
the algebra is commutative but nonassociative.

A physically reasonable choice of the damping factor $D(j)$ would be 
to mimic the Gaussian damping behavior of the flat coordinates in (\ref{eq:alginx}). So let me determine $D(j)$ by 
\begin{equation}
\exp\left[\beta \cos(\theta)\right]=\sum_{j,m}  D(j) Y_j^m(\theta,\varphi)Y_{j}^{-m}(0,0),
\label{eq:expansion}
\end{equation}
where $Y_j^m(\theta,\varphi)$ are the spherical harmonics, and  $\theta,\varphi$ are the angle coordinates 
on a two-sphere. 
Here the left-hand side is a Gaussian-like damping function on a two-sphere,
since $\exp\left[\beta \cos(\theta)\right]\sim const. \exp(-\beta \theta^2/2)$ at
$\theta\sim 0$.
Since the left-hand side does not depend on $\varphi$, only the terms with $m=0$ contribute in 
the right-hand side. By using some identities of the spherical harmonics, one obtains
\begin{equation}
D(j)=const. \int_{-1}^1 dz\, e^{\beta z}\, P_{j}(z)=const. (-1)^{j+\frac12} I_{j+\frac12}(-\beta),
\label{eq:damping}
\end{equation}
where $P_j$ and $I_j$ are the Legendre and the modified Bessel functions, respectively.
By putting (\ref{eq:damping}) into (\ref{eq:spherealg}) and computing (\ref{eq:scalaraction}),
one can numerically check that the action behaves at low $j$ as
\begin{equation}
S_{sphere}=(c_o+c_1 j(j+1)+\cdots)\psi_{(j,m)}^* \psi_{(j,m)}, 
\end{equation}
where $c_i$ are numerical constants.
Therefore, at low $j$, the scalar field action (\ref{eq:scalaraction}) reproduces the standard 
Laplacian on a two-sphere, when a fuzzy two-sphere is taken as a background.

\section{Summary and future prospects}
The rank-three tensor models may be interpreted as models for dynamical fuzzy spaces.    
The generalized Hermiticity condition on the rank-three tensor, which is the only dynamical
variable of the rank-three tensor models, corresponds to a cyclic property of the function algebras of the fuzzy spaces.
This paper has shown that, essentially due to this cyclic property, the fuzzy spaces have various physically interesting properties.
(i) Although the function algebras of the kind are nonassociative in general, various properties analogous
to quantum mechanics hold on the fuzzy spaces. (ii) The symmetry of the rank-three tensor models can be shown to be 
represented systematically by $n$-ary transformations on the fuzzy spaces. The transformations contain, for instance, 
diffeomorphism on the fuzzy spaces.
(iii) There exists a systematic procedure of truncating the function algebras of the kind, and it can be used to 
consider subspaces, compactifications, lattice theories, and coarse-graining procedures of fuzzy spaces 
in physical applications.

The discussions in this paper have been intended to be as general as possible without specifying physical problems.
Therefore the general implications obtained in this paper are expected to find wide applications in the future study of the 
rank-three tensor models and the associated fuzzy spaces.


\begin{thebibliography}{99}

\bibitem{Ambjorn:1990ge}
  J.~Ambjorn, B.~Durhuus and T.~Jonsson,
  ``Three-Dimensional Simplicial Quantum Gravity And Generalized Matrix
  Models,''
  Mod.\ Phys.\ Lett.\ A {\bf 6}, 1133 (1991).

\bibitem{Sasakura:1990fs}
  N.~Sasakura,
  ``Tensor Model For Gravity And Orientability Of Manifold,''
  Mod.\ Phys.\ Lett.\ A {\bf 6}, 2613 (1991).

\bibitem{Godfrey:1990dt}
  N.~Godfrey and M.~Gross,
  ``Simplicial Quantum Gravity In More Than Two-Dimensions,''
  Phys.\ Rev.\ D {\bf 43}, 1749 (1991).
  
\bibitem{Gurau:2011xp} 
  R.~Gurau and J.~P.~Ryan,
  ``Colored Tensor Models - a review,''  arXiv:1109.4812 [hep-th].  
  
\bibitem{Oriti:2011jm} 
  D.~Oriti,
  ``The microscopic dynamics of quantum space as a group field theory,''  arXiv:1110.5606 [hep-th].  
 

\bibitem{Sasakura:2005js} 
  N.~Sasakura,
  ``An Invariant approach to dynamical fuzzy spaces with a three-index variable,''  Mod.\ Phys.\ Lett.\ A {\bf 21}, 1017 (2006)  [hep-th/0506192].  

\bibitem{Sasakura:2011ma} 
  N.~Sasakura,
  ``Tensor models and 3-ary algebras,'' J.\ Math.\ Phys. {\bf 52}, 103510 (2011) [arXiv:1104.1463 [hep-th]].  


\bibitem{Sasakura:2006pq} 
  N.~Sasakura,
  ``Tensor model and dynamical generation of commutative nonassociative fuzzy spaces,''  Class.\ Quant.\ Grav.\  {\bf 23}, 5397 (2006)  [hep-th/0606066].  


\bibitem{Sasakura:2008pe} 
  N.~Sasakura,
  ``Emergent general relativity on fuzzy spaces from tensor models,''  Prog.\ Theor.\ Phys.\  {\bf 119}, 1029 (2008)  [arXiv:0803.1717 [gr-qc]].  

\bibitem{Sasakura:2009hs} 
  N.~Sasakura,
  ``Gauge fixing in the tensor model and emergence of local gauge symmetries,''  Prog.\ Theor.\ Phys.\  {\bf 122}, 309 (2009)  [arXiv:0904.0046 [hep-th]].  

\bibitem{Connes:1994yd} 
  A.~Connes,
  ``Noncommutative geometry,''  

\bibitem{Madore:1991bw}
  J.~Madore,
  ``The Fuzzy sphere,''
  Class.\ Quant.\ Grav.\  {\bf 9}, 69-88 (1992).


\bibitem{deMedeiros:2004wb}
  P.~de Medeiros and S.~Ramgoolam,
  ``Non-associative gauge theory and higher spin interactions,''
  JHEP {\bf 0503}, 072 (2005)
  [arXiv:hep-th/0412027].

\bibitem{Ramgoolam:2003cs}
  S.~Ramgoolam,
  ``Towards gauge theory for a class of commutative and nonassociative fuzzy
  spaces,''
  JHEP {\bf 0403}, 034 (2004)
  [arXiv:hep-th/0310153].

\bibitem{Ramgoolam:2001zx}
  S.~Ramgoolam,
  ``On spherical harmonics for fuzzy spheres in diverse dimensions,''
  Nucl.\ Phys.\  B {\bf 610}, 461 (2001)
  [arXiv:hep-th/0105006].
  
\bibitem{Sasai:2006ua}
  Y.~Sasai, N.~Sasakura,
  ``One-loop unitarity of scalar field theories on Poincare invariant commutative nonassociative spacetimes,''
  JHEP {\bf 0609}, 046 (2006).
  [hep-th/0604194].


\bibitem{Okubo:1990nv}
  S.~Okubo,
  ``Introduction to octonion and other nonassociative algebras in physics,''
  Cambridge, UK: Univ. Pr. (1995) 136 p. (Montroll memorial lecture series in mathematical physics, 2). 



\bibitem{Sasakura:2010rb} 
  N.~Sasakura,
  ``A Renormalization procedure for tensor models and scalar-tensor theories of gravity,''  Int.\ J.\ Mod.\ Phys.\ A {\bf 25}, 4475 (2010)  [arXiv:1005.3088 [hep-th]].  

\bibitem{FigueroaO'Farrill:2008bd}
  J.~M.~Figueroa-O'Farrill,
  ``Three lectures on 3-algebras,''
  [arXiv:0812.2865 [hep-th]].

\bibitem{deAzcarraga:2010mr}
  J.~A.~de Azcarraga and J.~M.~Izquierdo,
  ``n-ary algebras: A Review with applications,''
  J.\ Phys.\ A  {\bf 43}, 293001 (2010)
  [arXiv:1005.1028 [math-ph]].

\bibitem{Sasakura:2011nj} 
  N.~Sasakura,
  ``Tensor models and hierarchy of n-ary algebras,''  Int.\ J.\ Mod.\ Phys.\ A {\bf 26}, 3249 (2011)  [arXiv:1104.5312 [hep-th]].  

\bibitem{Sasakura:2011qg} 
  N.~Sasakura,
  ``Super tensor models, super fuzzy spaces and super n-ary transformations,''  Int.\ J.\ Mod.\ Phys.\ A {\bf 26}, 4203 (2011)  [arXiv:1106.0379 [hep-th]].  

\bibitem{Ferrari:1971at}
  R.~Ferrari and L.~E.~Picasso,
  ``Spontaneous breakdown in quantum electrodynamics,''
  Nucl.\ Phys.\  B {\bf 31}, 316 (1971).
  
\bibitem{Brandt:1974jw}
  R.~A.~Brandt and W.~C.~Ng,
  ``Gauge Invariance And Mass,''
  Phys.\ Rev.\  D {\bf 10}, 4198 (1974).
  
\bibitem{Borisov:1974bn}
  A.~B.~Borisov and V.~I.~Ogievetsky,
  ``Theory Of Dynamical Affine And Conformal Symmetries As Gravity Theory  Of
  The Gravitational Field,''
  Theor.\ Math.\ Phys.\  {\bf 21}, 1179 (1975)
  [Teor.\ Mat.\ Fiz.\  {\bf 21}, 329 (1974)].
  
\end{thebibliography}
\end{document}